\makeatletter
\newcommand{\dontusepackage}[2][]{%
  \@namedef{ver@#2.sty}{9999/12/31}%
  \@namedef{opt@#2.sty}{#1}}
\makeatother
\dontusepackage{subfigure}

\documentclass[]{article}

\usepackage{lmodern}
\usepackage{amssymb,amsmath}
\usepackage{ifxetex,ifluatex}
\usepackage[usenames,dvipsnames]{color}
\usepackage{fixltx2e} 
\ifnum 0\ifxetex 1\fi\ifluatex 1\fi=0 
  \usepackage[T1]{fontenc}
  \usepackage[utf8]{inputenc}
\else 
  \ifxetex
    \usepackage{mathspec}
    \usepackage{xltxtra,xunicode}
  \else
    \usepackage{fontspec}
  \fi
  \defaultfontfeatures{Mapping=tex-text,Scale=MatchLowercase}
  
\fi
\IfFileExists{upquote.sty}{\usepackage{upquote}}{}
\IfFileExists{microtype.sty}{%
\usepackage{microtype}
\UseMicrotypeSet[protrusion]{basicmath} 
}{}
\usepackage[margin=1.0in,bottom=1.5in]{geometry}
\usepackage[square,numbers,sort&compress]{natbib}
\usepackage{listings}
\usepackage{graphicx}
\makeatletter
\def\maxwidth{\ifdim\Gin@nat@width>\linewidth\linewidth\else\Gin@nat@width\fi}
\def\maxheight{\ifdim\Gin@nat@height>\textheight\textheight\else\Gin@nat@height\fi}
\makeatother
\setkeys{Gin}{width=\maxwidth,height=\maxheight,keepaspectratio}
\usepackage{caption}
\usepackage{float}
\usepackage{subfig}
\ifxetex
  \usepackage[setpagesize=false, 
              unicode=false, 
              xetex]{hyperref}
\else
  \usepackage[unicode=true]{hyperref}
\fi
\hypersetup{breaklinks=true,
            bookmarks=true,
            pdfauthor={},
            pdftitle={Accelerating innovation with software abstractions for scalable computational geophysics},
            colorlinks=true,
            citecolor=black,
            urlcolor=blue,
            linkcolor=black,
            pdfborder={0 0 0}}
\usepackage[preprint]{neurips_2019}
\usepackage{url}
\usepackage{booktabs}
\usepackage{amsfonts}
\usepackage{nicefrac}
\usepackage{microtype}

\title{Accelerating innovation with software abstractions for scalable
computational geophysics}
\author{Mathias Louboutin\textsuperscript{*,1}, Philipp
Witte\textsuperscript{2}, Ali Siahkoohi\textsuperscript{1}, Gabrio
Rizzuti\textsuperscript{3}, \\ \textbf{Ziyi Yin}\textsuperscript{1}, \textbf{Rafael
Orozco}\textsuperscript{1} \textbf{and Felix J.
Herrmann}\textsuperscript{1}\\\textsuperscript{1} Georgia Institute of
Technology, \textsuperscript{2} Microsoft Research, \textsuperscript{3}
Utrecht University}
\date{}

\begin{document}
\maketitle

\section{Summary}\label{summary}

We present the \href{https://github.com/slimgroup}{SLIM} open-source
software framework for computational geophysics, and more generally,
inverse problems based on the wave-equation (e.g., medical ultrasound).
We developed a software environment aimed at scalable research and
development by designing multiple layers of abstractions. This
environment allows the researchers to easily formulate their problem in
an abstract fashion, while still being able to exploit the latest
developments in high-performance computing. We illustrate and
demonstrate the benefits of our software design on many geophysical
applications, including seismic inversion and physics-informed machine
learning for geophysics(e.g., loop unrolled imaging, uncertainty
quantification), all while facilitating the integration of external
software.

\vspace*{-0.5cm}

\section{Introduction}\label{introduction}

\vspace*{-0.25cm}

Software development for exploration geophysics has been traditionally
driven by performance. Though this has resulted in very efficient code,
the usability and portability of these codes has been compromised as a
result of the absence of user-level design. Abstractions and user
interfaces at a high level have been developed in recent decades to
facilitate research and development. A number of such interfaces are
available, ranging from programming languages, e.g., Python
\citep{python}, Julia \citep{julia}, and Matlab, to domain-specific
languages (DSLs), including RVL \citep{ricevector}, Firedrake
\citep{firedrake}, and Devito \citep{devito-api, devito-compiler}.
Additionally, there has been a community initiative towards open-source
codes and reproducibility, led by Madagascar \citep{Madagascar}. These
efforts have brought attention to the need for user abstraction to
easily translate mathematics into code without having to manually
refactor thousands of lines of code.

Motivated by this background, we introduce our fully open-source
software \href{https://github.com/slimgroup}{SLIM} framework based on
high-level abstractions and separation of concerns. Our software design
relies on three principles: (1) A high-level abstraction that represents
the mathematical problem at hand; (2) Scalability with vertical
integration of high-performance software and DSLs; (3) Inter-operability
through the adoption of language standards (object oriented, multiple
dispatch, inheritence, \ldots{}). In the following, we will detail these
three points, highlighting their importance and implementation. We will
follow with some illustrations of the integration with external software
to demonstrate how research and development can be eased and potentially
accelerated by these design principles.

\vspace*{-0.5cm}

\section{Software design}\label{software-design}

\vspace*{-0.25cm}

The fundamental objective of a scientific software framework is to
provide users with an interface that allows them to express their own
scientific problems. In particular, the interface should provide
abstractions to define the problem as closely as possible to its
mathematical formulation. With this design philosophy, research and
development turnaround time can be drastically reduced with quick
prototyping and testing of new ideas and algorithms. Our software
framework is grounded in this idea and led to the development of legacy
software \citep{lin2015IIPFWIsdh, dasilva2017uls} adopted by industry
for real-world application. Based on these ideas and modern programming
paradigms, we designed a high-level framework that encapsulates the
mathematical definition of geoscientific problems at every level. These
different levels are handled through the vertical integration of
domain-specific languages (DSLs). At the lowest level,
\href{https://github.com/devitocodes/devito}{Devito} provides a symbolic
DSL for the definition of wave-equation and a just-in-time compiler to
target the available hardware with near-optimal performance. On top of
\href{https://github.com/devitocodes/devito}{Devito}, we developed
\href{https://github.com/slimgroup/JUDI.jl}{JUDI}, a linear algebra DSL
for wave-equation based modeling and inversion. Finally, we created a
machine learning framework that allows us to integrate
\href{https://github.com/devitocodes/devito}{Devito} propagators and
\href{https://github.com/slimgroup/JUDI.jl}{JUDI} Linear Operators into
machine learning (ML) frameworks (PyTorch and Flux) opening the door to
ML-augmented geophysical inverse problems and uncertainty quantification
(UQ). Finally, throughout these different layers, we committed to follow
language standards allowing easy interfacing and integration with
external software and frameworks.

\vspace*{-0.5cm}

\section{Wave-equation based
inversion}\label{wave-equation-based-inversion}

\vspace*{-0.25cm}

Linear operators are at the core of applied geophysics. Some relevant
examples include data filtering tools such as Fourier-based f-k filters,
the Radon transform, NMO corrections to traditional imaging operator
such as Kirchhoff migration, post-stack migration and wave-equation
based inversion where the discrete wave-equation is a linear operator.
One of the first frameworks for abstract matrix-free linear algebra was
sPOT\citep{friedlander2009SCAIMspot}, later extended to wave-equation
based inversion and distributed computing for wavefield reconstruction
\citep{kumar2010SINBADpoe}. While its adoption was limited by its
implementation in Matlab, such user oriented abstraction laid the ground
for modern abstracted frameworks such as \texttt{JOLI}\citep{JOLI} in
Julia and \texttt{pyLops} in Python \citep{RAVASI2020100361}.

Similarly, wave-equation based inversion can be trivially formulated as
simple least-squares optimization problem for the non-linear
wave-equation represented as an abstract matrix $\mathbf{A}(\mathbf{m})$
parametrized by the model parameter $\mathbf{m}$:
\begin{equation*}
\begin{split}
    \underset{\mathbf{m}}{\operatorname{minimize}} \hspace{0.2cm} \Phi(\mathbf{m}) = \sum_{i=1}^{n_s} ||\mathbf{P}_r \mathbf{A}(\mathbf{m})^{-1}\mathbf{P}_s^{\top} \mathbf{q}_i - \mathbf{d}_i ||^2_2, \\
\end{split}
\end{equation*}
 Building on modern software solutions
\href{https://github.com/slimgroup/JUDI.jl}{JUDI}
\citep{witte2018alf, mathias_louboutin_2022_5893940} provides a
high-level interface allowing to define a solver for this problem with a
few lines of code. For example, FWI using Gauss-Newton updates at each
iteration can be summarized in as little as five lines
(Listing~\ref{fig:fwi-gn}).

\begin{lstlisting}[caption={FWI with Gauss-Newton updates using
\href{https://github.com/slimgroup/JUDI.jl}{JUDI}. The complete example
is available and reproducible in the
\href{https://github.com/slimgroup/JUDI.jl}{JUDI} github repository.
\vspace*{-0.65cm}}, label=fig:fwi-gn, float=htbp]
# Gauss-Newton method
F = judiModeling(model, srcGeom, recGeom)
J = judiJacobain(F, q)
for j=1:maxiter
    d_pred = F*q
    fhistory_GN[j] = .5f0*norm(d_pred - d_obs)^2
    # Gauss-Newton update
    p = lsqr(J, d_pred - d_obs)
    model0.m .= proj(model0.m .- reshape(p, model0.n))
end
\end{lstlisting}

\begin{figure}
\centering
\includegraphics[width=0.500\hsize]{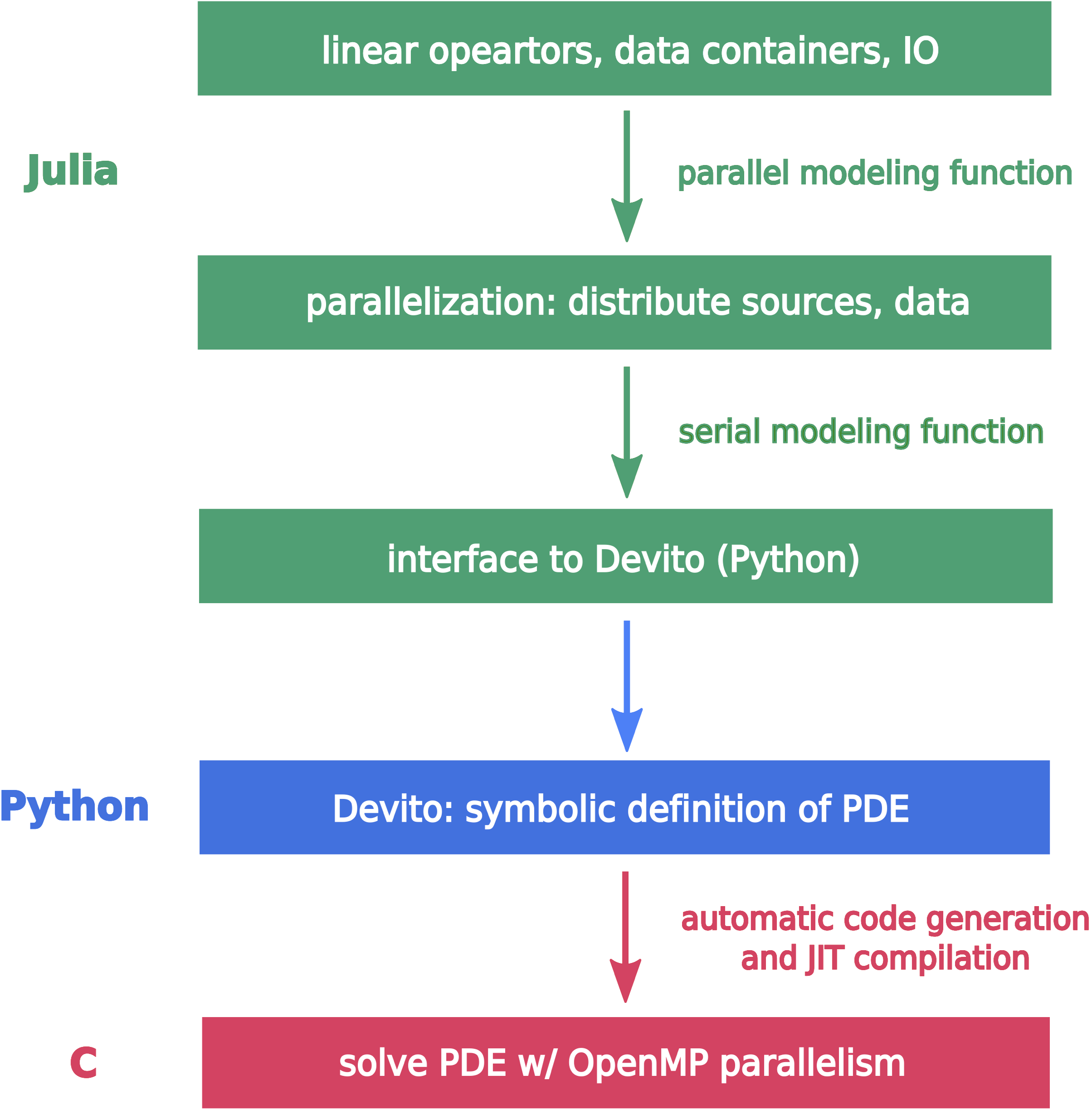}
\caption{Schematics of the vertical integration in
\href{https://github.com/slimgroup/JUDI.jl}{JUDI} that enables at scale
inversion through layers of abstractions.}\label{fig:vert}
\end{figure}

\begin{figure}
\centering
\includegraphics[width=1.000\hsize]{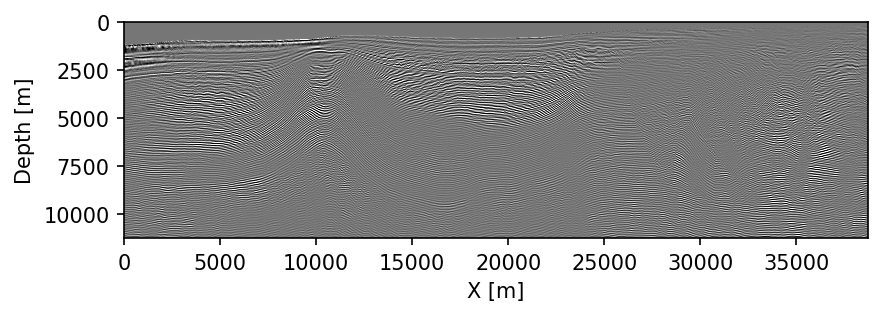}
\caption{2D RTM on the 2007 BP TTI model with a marine acquisition. This
RTM was run on a GPU without moving off to the CPU at any time using
randomized trace estimation as an extension of
\href{https://github.com/slimgroup/JUDI.jl}{JUDI}
\citep{louboutin2021SEGulm, mathias_louboutin_2021_5605055}.
\vspace*{-0.5cm}}\label{fig:2d-bp}
\end{figure}

In addition to offering a high-level, mathematical interface to the
wave-equation and related linear operators,
\href{https://github.com/slimgroup/JUDI.jl}{JUDI} builds on multiple
layers of abstractions while still providing computational performance
that is comparable to the state-of-the-art\citep{devito-compiler}.
\href{https://github.com/slimgroup/JUDI.jl}{JUDI} integrates
\href{https://github.com/devitocodes/devito}{Devito} as a backend for
the actual wave-equation solves. Devito is a stencil DSL
\citep{devito-api, devito-compiler} that offers a symbolic user-level
interface to the underlying numerical solver. This symbolic interface
allows the user to easily and mathematically define generic partial
differential equations (PDEs), such as the acoustic,
tilted-transverse-isotropic (TTI), or elastic wave-equations. The
underlying software then provides a code generation framework than can
generate highly-performant code for numerous architectures, (CPU, GPUs,
ARM, POWER). Its ease of use and high-level interface have contributed
to its adoption in the oil-and-gas industry for production and research
at scale \citep{cofii}. We summarize the vertical integration of the
different technologies (compiler,
\href{https://github.com/devitocodes/devito}{Devito},
\href{https://github.com/slimgroup/JUDI.jl}{JUDI}, \ldots{}) in
Figure~\ref{fig:vert}, which highlights the different levels of
abstraction. The 2007 BP TTI dataset, for example, can be setup and
processed with RTM within days. The results are shown in
Figure~\ref{fig:2d-bp}.

\subsection{Interoperability}\label{interoperability}

While a proper separation of concerns can lead to highly usable,
portable, and performant software, another important aspect to consider
is interoperability, namely the ability to integrate software from
different sources and organizations. We now describe how the design of
our software framework allows for easy interfacing with external
frameworks. One major drawback of proprietary and low-level software is
the limited capability to interface and interact with external software.
Through open-source code development, and committing to
language-specific standards (such as proper usage of types, hierarchy,
inheritance, etc.) in Julia and Python, we enable seamless
interoperability with external software and frameworks. This greatly
increases the ability to perform research, which involves being able to
easily test and combine different ideas. For instance, we combine core
Julia packages, our numerical optimization toolbox
\href{https://github.com/slimgroup/SlimOptim.jl}{SlimOptim.jl}
\citep{mathias_louboutin_2021_5196691}, a constrained optimization
software
\href{https://github.com/slimgroupSetIntersectionProjection.jl}{SetIntersectionProjection.jl}
\citep{peters2019algorithms}, and COFII's wave-equation propagators
\citep{cofii}, in order to setup a full waveform inversion (FWI)
exercise. We were able to perform FWI, in parallel, on the Marmousi-II
model by taking advantage of the best technology available.
Additionally, this demonstrative example, and in general
\href{https://github.com/slimgroup/JUDI.jl}{JUDI} (or COFII), can
trivially be deployed in the cloud using once again dedicated software
abstractions \citep{cofii, mathias_louboutin_2022_6386831}.

\begin{lstlisting}[caption={Projected Quasi-Newton FWI with multiple
software packages. The complete example can be found
\href{https://github.com/slimgroup/ConstrainedFWIExamples/blob/master/notebooks/02_constr_fwi_jetpack.ipynb}{here}
\vspace*{-0.65cm}}, label=fig:inter, float=htbp]
function objective(F, velocity, d_obs)
    J = jacobian(F, velocity)
    d_pred = F * velocity # Forward modeling with COFFI
    G = J' * ({d_pred.- d_obs) # Gradient with COFII
    f = .5f0*norm(d_pred .- d_obs)^2 # Misfit
    return f, G
end
# Projection on TV+bounds, SetIntersectionProjection.jl
prj = setup_projection(...)
# Projected Quasi-Newton (l-BFGS) with SlimOptim.jl
sol = pqn(x->objective(F, x, d_obs), vec(v2), prj);
\end{lstlisting}

Since different software modules can be pieced together with minimal
effort, it greatly aids the development and testing of new research.
This flexibility allows, for instance, the integration of machine
learning into conventional algorithm pipelines in a plug-and-play
fashion. In the following section, we provide some notable examples of
this approach.

\vspace*{-0.5cm}

\section{Machine learning}\label{machine-learning}

\vspace*{-0.25cm}

Thanks to recent theoretical and practical advances, deep learning has
seen a wide adoption in various geophysical applications ranging from
processing (e.g., denoising, multiple removal) to inverse problems and
seismic interpretation
\citep[e.g.,][]{zhang2018deep, wang2019deep, yang2019deep}. One of the
core challenges in the adoption of deep learning for seismic
applications is the integration of existing codes (e.g.~seismic
propagators) into deep-learning frameworks based on automatic
differentiation (AD). To be able to differentiate through networks that
contain both standard PyTorch/Tensorflow layers, as well as third party
functions, such as a forward modeling kernel, the third party code must
have manually implemented gradients and must be properly integrated into
the deep learning framework. Our high-level linear algebra abstraction
framework, which sits on top of our Devito-based propagators,
facilitates this integration, as deep learning frameworks like PyTorch
or Flux are already able to backpropagate through a linear operator
(namely by applying its adjoint to the data residual). This allows us to
implement deep neural networks in Flux, in which we can combine standard
deep learning layers with external third party functions,
e.g.~convolutional layers with migration/demigration operators. This
capability has enabled various research projects in our group, including
surface-related multiple elimination \citep{siahkoohi2019SEGsrm},
dispersion attenuation \citep{siahkoohi2019itl}, and ML-augmented
imaging and uncertainty quantification
\citep{siahkoohi2019TRnna, siahkoohi2020EAGEdlb}. We show in the
following two examples of machine learning for exploration geophysics
that take advantage of these high-level abstractions to interface PDE
solvers (\href{https://github.com/slimgroup/JUDI.jl}{JUDI},
\href{https://github.com/devitocodes/devito}{Devito}) with deep learning
frameworks.

\subsection{Seismic imaging with deep priors and uncertainty
quantification}\label{seismic-imaging-with-deep-priors-and-uncertainty-quantification}

\begin{figure*}
\centering
\subfloat[\label{fig:mle-deep-prior-1}]{\includegraphics[width=0.300\hsize]{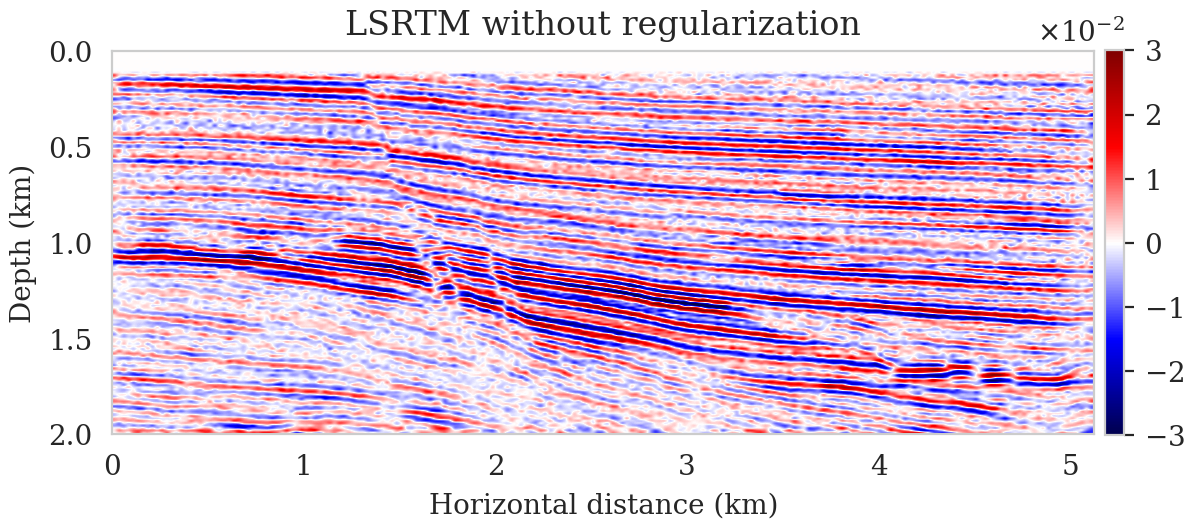}}
\subfloat[\label{fig:cm-deep-prior-1}]{\includegraphics[width=0.300\hsize]{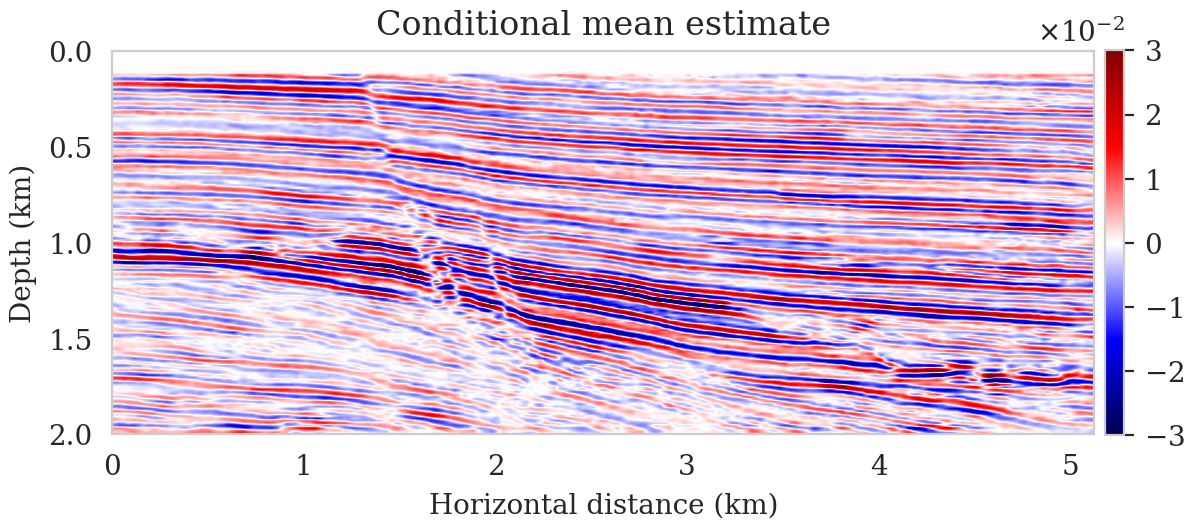}}
\subfloat[\label{fig:std-deep-prior-1}]{\includegraphics[width=0.300\hsize]{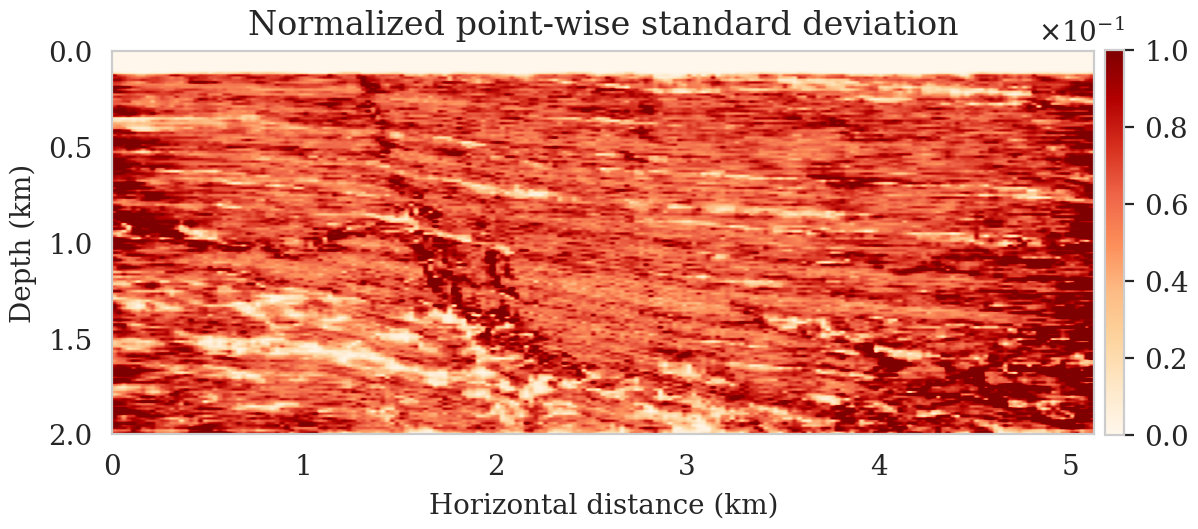}}
\caption{Imaging a 2D subset of the
\href{https://wiki.seg.org/wiki/Parihaka-3D}{Parihaka}
\citep{WesternGeco2012} dataset with deep priors
\citep{siahkoohi2021deep}. (a) LSRTM without any regularization. (b) The
conditional (posterior) mean estimate, using the deep prior as
regularization. (c) Normalized pointwise standard
deviation.}\label{fig:deep-prior-1}
\end{figure*}

Since we can integrate
\href{https://github.com/devitocodes/devito}{Devito}'s highly optimized
wave-equation solvers with the PyTorch deep learning library, we propose
to use deep priors \citep{Lempitsky} to regularize seismic imaging,
where we reparameterize the seismic image as the output of an untrained
convolutional neural network (CNN). This approach acts as a
regularization in the image space
\citep{Lempitsky, Cheng_2019_CVPR, dittmer2020regularization}, which
exploits the inductive bias of the CNN in representing images without
noisy artifacts \citep{Lempitsky}. We perform Bayesian inference via a
gradient-based Markov chain Monte Carlo (MCMC) algorithm
\citep{welling2011bayesian}, where each iteration requires
differentiating the action of the linearized Born scattering operator on
the CNN output with respect to the CNN's weights. In order to have
access to the automatic differentiation utilities of PyTorch, we expose
\href{https://github.com/devitocodes/devito}{Devito}'s matrix-free
implementations for the migration operator and its adjoint to PyTorch
via
\href{https://github.com/slimgroup/Devito4Pytorch.jl}{Devito4PyTorch}
\citep{ali_siahkoohi_2021_4610087}. This is exemplified in
Listing~\ref{lst:deepprior}. Figure~\ref{fig:deep-prior-1} summarizes
the results of seismic imaging and uncertainty quantification with this
approach, which are borrowed from \citet{siahkoohi2021deep}.

\begin{lstlisting}[caption={Seismic imaging and uncertainty
quantification with PyTorch and
\href{https://github.com/devitocodes/devito}{Devito} as a wave-equation
solver.
\vspace*{-0.65cm}}, language=Python, label=lst:deepprior, float=htbp]
def deep_prior_imaging(d_obs, n, maxiter=1000):
    # PyTorch Integrated Devito Jacobian.
    J = ForwardBornLayer(model, geometry)
    # Initialize deep prior CNN with random input.
    G = deep_prior_net()
    z = torch.randn([1, 1, n[1], n[2]])
    # Setup pSGLD MCMC sampler.
    optim, samples = pSGLD(G.parameters(), lr=0.001), []
    for i in range(maxitr):
        # Compute predicted perturbation model and data.
        x = G(z)
        d_pred = J(x)
        # Compute the negative log-posterior.
        nlp = n_log_posterior(d_pred, d_obs, G.parameters())
        # Compute the gradient w.r.t. the CNN weights,
        nlp.backward()
        # Sample with pSGLD.
        optim.step()
        samples.append(x.detach().numpy())
    return samples
\end{lstlisting}

\subsection{Loop-unrolled seismic
imaging}\label{loop-unrolled-seismic-imaging}

\begin{figure}
\centering
\includegraphics[width=0.600\hsize]{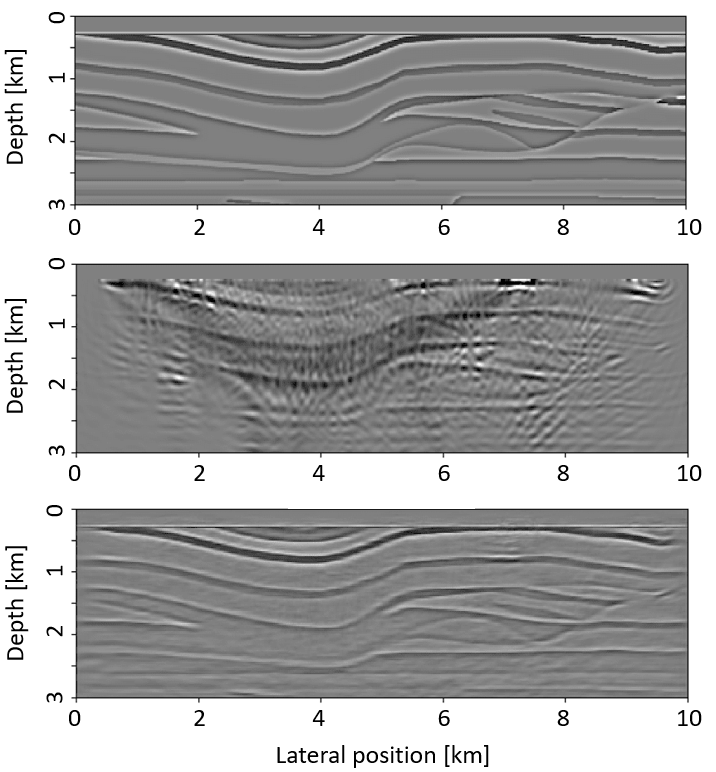}
\caption{Loop-unrolled LSRTM on a test 2D slice of the 2D overthrust
model. (a) True image, (b) RTM of simultaneous shot and (c)
loop-unrolling of.}\label{fig:lulsrtm}
\end{figure}

Another set of applications that are enabled by the proper integration
of abstract seismic modeling operators into deep learning frameworks are
loop-unrolled optimization algorithms such as the learned primal-dual
reconstruction \citep{andrychowicz2016learning, adler2018learned}. These
are special types of networks that follows the general structure of
gradient-based optimization algorithms, in which conventional gradients
are augmented by additional neural network layers. Using our integration
of \href{https://github.com/slimgroup/JUDI.jl}{JUDI} operators into Flux
via the \href{https://github.com/slimgroup/JUDI4Flux.jl}{JUDI4Flux}
package \citep{philipp_a_witte_2019_4301018}, we apply the loop-unrolled
network architecture from \citep{adler2017solving} to seismic imaging.
Every iteration of the loop-unrolled algorithm consists of computing the
(conventional) gradient of the LS-RTM objective function $g$, using the
forward and adjoint linearized Born scattering operator, followed by the
application of a shallow CNN (Listing~\ref{lst:unroll}). The network
takes a single simultaneous (super-) shot record as the input and
predicts an LS-RTM (or true) image. We train the network in a supervised
fashion, in which we minimize the misfit between the predicted image and
the true image (whereas in reality, one would train with available
LS-RTM images). Figure~\ref{fig:lulsrtm} shows the output of the
loop-unrolled gradient descent algorithm, Listing~\ref{lst:unroll},
after training the network for 2000 iterations on a training dataset of
2,000 data-image pairs. The example is not meant to represent a
realistic scenario (as the true images which were used in the training
process are obviously unknown), but serve as a proof of concept on how
to augment physical models with data-driven approaches. The complete
example and additional variations are available at
\href{https://github.com/slimgroup/LoopUnrolledSeismicImaging}{LoopUnrolledSeismicImaging}.

\begin{lstlisting}[caption={Example of a physics-augmented neural
network for seismic imaging. The network consists of 10 iterations of a
loop-unrolled gradient descent algorithm, in which the conventional
LS-RTM gradient $g$ is augmented through convolution layers. The input
into the network is the observed seismic data and the output is the
predicted image. \vspace*{-0.65cm}}, label=lst:unroll, float=htbp]
function loop_unrolled_lsrtm(d_obs, n; maxiter=10)
    x = randn(Float32, n[1], n[2], 1, 1)
    nx, ny, nc, nb = size(x)
    s = randn(Float32, nx, ny, 5, nb) # memory term
    for j=1:maxiter
        g = adjoint(J)*(J*vec(x) - d_obs)
        g = reshape(Flux.normalise(g), nx, ny, 1, 1)
        u = cat(x, g, s, dims=3)
        u = bn1(conv1(u)); u = bn2(conv2(u)); u = bn3(conv3(u));
        s = relu.(u[:, :, 2:6, :])
        dx = u[:, :, 1:1, :]
        x += dx
    end
    return vec(x)
end
\end{lstlisting}

\subsection{End-to-end inversion for geological carbon storage
monitoring}\label{end-to-end-inversion-for-geological-carbon-storage-monitoring}

\begin{figure}
\centering
\includegraphics[width=1.000\hsize]{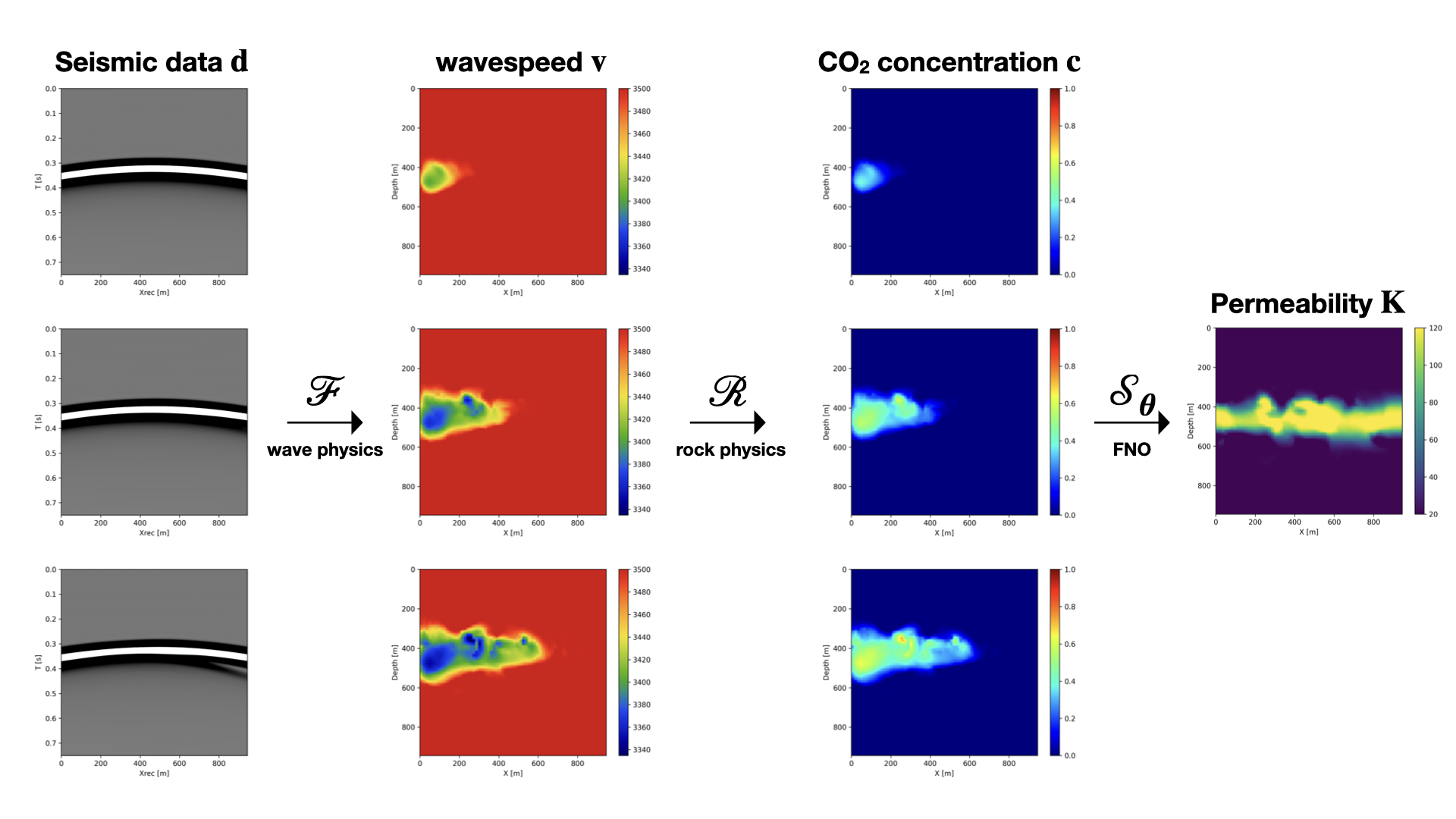}
\caption{End-to-end inversion of CO\textsubscript{2} concentration from
seismic measurements. In order to obtain this end-to-end result, the
three operators representing the fluid-flow modeling, rock property
modeling, wave modeling, and their derivatives are integrated by
combining AD and manually defined gradients.}\label{fig:end-to-end}
\end{figure}

Finally, we show that we can easily build a framework for seismic
monitoring of geological carbon storage that integrates PDE solvers,
deep learning models and physical constraints. These monitoring problems
rely on three different types of physics \citep{li2020coupled}:
fluid-flow physics, rock physics and wave physics. Given time-lapse
seismic data collected over multiple years, we jointly invert for the
rocks intrinsic permeability which can be used to recover the
CO\textsubscript{2} concentration. By leveraging our high-level
abstractions and AD, we can directly differentiate through all three
physics solvers which map permeability to seismic data and minimize a
fully coupled data misfit objective function. At each iteration, given
an estimate of the permeability $K$, we model seismic data where the
subsurface velocity is translated from the solution of the fluid-flow
simulation. We can then compute the standard data misfit and use
automatic differentiation to compute the permeability update. Because
each step is abstracted, we can easily swap each physical solver with a
different one, such as replacing the fluid-flow solver by a trained
Fourier Neural Operator (FNO) for computational efficiency
\citep{li2020fourier}. We show in Figure~\ref{fig:end-to-end} that we
can recover the permeability from seismic measurements using a
pre-trained FNO in place of a fluid-flow solver.

\begin{lstlisting}[caption={Example of an end-to-end coupled inversion
for seismic monitoring of geological carbon storage. A pre-trained FNO
$S$ is used as a surrogate for fluid-flow simulation. In each iteration,
we calculating the seismic data misfit, compute the gradient with
respect to input permeability, $K$, via AD, and update it according to
an optimizer $opt$. \vspace*{-0.65cm}}, label=lst:seis4ccs, float=htbp]
opt = RMSprop()
for j=1:maxiter
    theta = params(K)
    grads = gradient(theta) do
        c = S(K); v = R(c); d_pred = F(v)
        return 0.5f0 * norm(d_pred-d_obs)^2f0
    end
    for p in theta
        update!(opt, p, grads[p])
    end
end
\end{lstlisting}

\vspace*{-0.5cm}

\section{Conclusions}\label{conclusions}

\vspace*{-0.25cm}

Through this paper, we have introduced a software design philosophy
aimed at enabling research and development at scale in geoscience, with
the potential to generalize to any inverse problem such as medical
imaging. We demonstrated through carefully chosen example that
high-level abstractions allow to express complex problem in a clear and
representative way without incurring additional computational costs. Our
software framework already allows for a wide range of application,
including industry scale inversion and cutting-edge physics-informed
deep learning for geophysics. We intend to build additional capabilities
to tackle modern computing environment such as Cloud computing, task
dedicated accelerators (i.e TPUs) and non-convex optimization.

\vspace*{-0.5cm}

\section{Acknowledgment}\label{acknowledgment}

\vspace*{-0.25cm}

This research was carried out with the support of Georgia Research
Alliance and partners of the ML4Seismic Center. We thank J. Washbourne
at CVX for the fruitful discussions on open source software development
and interoperability.

\bibliography{louboutin_seg22}

\end{document}